\documentclass{optica-article}
\journal{opticajournal} % for journals or Optica Open
\articletype{Research Article}
\usepackage{lineno}
\usepackage{svg}
\begin{document}

\title{Zeptosecond-level timing-noise floor in optical-to-terahertz photomixing}
\author{William F. McGrew\authormark{1,*}, James Greenberg\authormark{1}, Scott C. Egbert\authormark{1}, Brendan M. Heffernan\authormark{1}, Y. Lange Simmons\authormark{1}, Tatsuya Tanigawa\authormark{1}, Kazutoshi Kato\authormark{2}, Tadao Nagatsuma\authormark{3}, Antoine Rolland\authormark{1}}
\address{\authormark{1}IMRA America, Inc, Boulder Research Labs, Longmont, CO 80501, USA\\
\authormark{2}Kyushu University, Graduate School of Information Science and Electrical Engineering, 744 Motooka Nishi-ku, Fukuoka, Japan\\
\authormark{3}The University of Tokyo, Institute for Photon Science and Technology, 7-3-1 Hongo, Bunkyo-ku, Tokyo Japan}
\email{\authormark{*}wmcgrew@imra.com} %% email address is required; see note below about the corresponding author designation
% use {asbstract*} to suppress the copyright line. Copyright information will be added in production
\begin{abstract*} %100 words... ish
Optical generation of terahertz (0.1 to 10\,THz) carriers can, in principle, preserve the spectral purity of an optical reference without the noise penalty associated with electronic frequency multiplication. Here, we identify and suppress the dominant excess-noise mechanisms that limit this conversion in high-power uni-travelling-carrier photodiodes. We characterize amplitude-to-phase conversion as functions of photocurrent and reverse bias and identify operating parameters to minimize the effect. We quantify the white phase noise introduced by optical amplification and show that this contribution can be eliminated by directly driving the photodiode with a high-power optical source. With the excess noise mitigated, we use cross-correlation phase-noise measurements to observe a white phase-noise floor of $-161.0$~dBc/Hz at 263.2~GHz, corresponding to a timing-noise floor of 7.6~zs/$\sqrt{\mathrm{Hz}}$. The remaining floor follows the photocurrent dependence expected from shot noise. Extending the measurement across the full WR-3.4 band, from 220 to 330~GHz, yields timing-noise floors below 10~zs/$\sqrt{\mathrm{Hz}}$ throughout, with a minimum of 6.0~zs/$\sqrt{\mathrm{Hz}}$. These results establish direct photomixing as a route to sub-10-zs/$\sqrt{\mathrm{Hz}}$ white timing-noise floors for continuous-wave sources at terahertz frequencies and identify the noise processes that must be controlled to reach this regime.
\end{abstract*}

%%%%%%%%%%%%%%%%%%%%%%%%%%  body  %%%%%%%%%%%%%%%%%%%%%%%%%%

\section{Introduction}

Terahertz (THz) applications that utilize phase coherence  are ultimately limited by the spectral purity of the carrier. In coherent wireless communications, oscillator phase noise limits the fidelity with which high-order modulation formats can be transmitted and recovered \cite{Proakis2008,Akyildiz2022,Maekawa2024}. In spectroscopy and precision measurement, the same fluctuations broaden narrow spectral features and limit the resolution of small frequency differences. Generating carriers with sufficiently low phase noise is therefore essential to the performance of systems across the 0.1 to 10\,THz range \cite{Wang2018, Greenberg2025, McGrew2025}.

A direct approach to THz wave generation is via an electronic oscillator with a native frequency at the target frequency. Resonant-tunneling-diode and transferred-electron oscillators generate THz carriers directly, but the resonators available at these frequencies have relatively low quality factors (Q). As a result, their free-running spectral purity remains far above the levels required for precision synthesis: direct electronic oscillators reported from roughly 80\,GHz into the THz range exhibit phase noise of order $-80$\,dBc/Hz at 1\,MHz offset \cite{Harvey1994, Kobayashi1999, Nakamura2015, Asada2021}. Further extrapolating their reported $1/f^2$ behavior, such sources remain above $-160$\,dBc/Hz out to Fourier frequencies of order GHz. Phase locking to a stable reference suppresses close-in fluctuations, but the noise outside the effective locking bandwidth remains set by the free-running device \cite{Ogino2017, Greenberg2026}. 

Limitations on resonator quality factor become much less severe at microwave frequencies below about 20\,GHz, where high-$Q$ resonators are readily and commercially available. This naturally motivates generating the carrier where spectral purity is more favorable and translating it to the desired terahertz frequency.

An experimentally common approach is frequency multiplication. Ideal multiplication by a factor $N$ theoretically increases phase noise by $20\log_{10}N$\,dB, while leaving the timing noise $S_x=S_\phi/(2\pi\nu_0)^2$ unchanged. Therefore, timing noise provides a natural basis for comparing sources operating at widely different carrier frequencies. Furthermore, the process of multiplication and amplification increases the noise floor as well, due to power losses and amplifier noise figure. The residual phase noise for a multiplication chain from 13 to 104 GHz has been measured to be -139 dBc/Hz at 1 MHz for a single synthesis chain \cite{Bara-Maillet2015}. This implies a minimum timing noise of 240\,zs/$\sqrt{\text{Hz}}$, even neglecting the phase noise of the microwave reference itself. A reported full 300-GHz reference based on electronic frequency multiplication -- to our knowledge, the lowest noise floor achieved for such a system -- has a noise floor of -121 dBc/Hz, amounting to 650 zs/$\sqrt{\text{Hz}}$ \cite{Wrana2026}.

Photonic sources provide a fundamentally different route to the same THz frequency regime. Two mutually coherent optical fields separated by the desired terahertz frequency can be photomixed directly in a high-speed photodiode, transferring their relative phase to the generated electrical carrier in a single conversion step and without electronic multiplication. Optical references such as dual-wavelength Brillouin lasers (DWBLs) and Fabry-Perot cavities have demonstrated exceptionally low relative phase noise, while optical frequency \textit{division} can reduce phase fluctuations still further, by the same quantity of $20 \log N$, before photodetection \cite{Tetsumoto2021, Shin2023, Heffernan2024, Chen2024, Groman2025, He2025, Egbert2026}. The optically generated THz (opto-THz) source employed in this work is the DWBL. By using a common fiber cavity, perturbations are shared between the two optical lines of the DWBL, allowing synthesis of ultra-stable THz waves. Once the optical reference is sufficiently quiet, the spectral purity of the generated carrier is set by the noise added during optical-to-electrical conversion. Determining that added-noise floor is consequently a fundamental limit to the performance of direct photomixing.

At terahertz frequencies, this conversion is commonly performed using uni-travelling-carrier photodiodes (UTC-PDs) \cite{Ishibashi1999}. Several mechanisms can add phase noise during photomixing and will be explored extensively in this work. Optical intensity fluctuations can be converted into phase fluctuations through the nonlinear response of a photodiode, optical amplification introduces white noise through amplified spontaneous emission, and photocurrent shot noise imposes a fundamental detection limit. These processes have been studied extensively in photonic microwave generation, where timing-noise floors as low as 20\,zs/$\sqrt{\mathrm{Hz}}$ have been reported at 8\,GHz \cite{Fortier2011,Xie2017,Kalubovilage2020, Kalubovilage2022}. Their impact in THz photomixing is less well established, particularly under the high-power, near-saturation conditions required for efficient generation. Recent UTC-PDs employing high-thermal-conductivity SiC substrates have increased the available output power at these frequencies to the milliwatt level \cite{Qian2024, Ssali2026}, reducing the relative importance of additive noise.

In this work, we identify and quantify the principal mechanisms that limit the white-noise floor of optical-to-terahertz photomixing. We first characterize amplitude-to-phase conversion in a high-power UTC-PD as a function of photocurrent and reverse bias, and identify operating conditions for which its contribution becomes negligible. We then measure the white phase noise introduced by optical amplification and eliminate this contribution by driving the photodiode directly with a high-power optical reference. Finally, by varying the photocurrent, we show that the remaining white-noise floor follows the scaling expected from photocurrent shot noise.

Using cross-correlation phase-noise measurements, we measure a white phase-noise floor of $-161.0$\,dBc/Hz at 263.2\,GHz, corresponding to a timing-noise floor of 7.6\,zs/$\sqrt{\mathrm{Hz}}$, more than an order of magnitude below reported buffered multiplication-chain values. Tuning the optical frequency difference across the WR-3.4 band, we demonstrate timing-noise floors below 10\,zs/$\sqrt{\mathrm{Hz}}$ from 220 to 330\,GHz, with values as low as 6.0\,zs/$\sqrt{\mathrm{Hz}}$.

\section{Noise processes in optical-to-THz photomixing}

\begin{figure}[t]
    \centering
    \includegraphics[width=\linewidth]{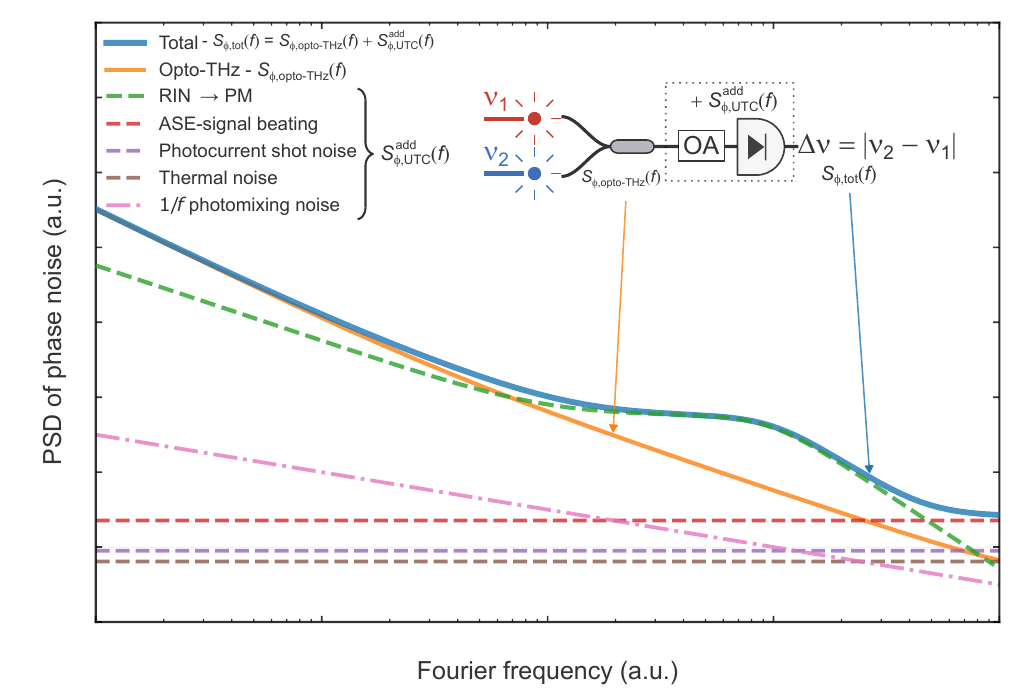}
    \caption{
    Phase-noise budget for direct optical-to-THz photomixing.
    Two mutually coherent optical fields separated by $\Delta\nu = |\nu_2-\nu_1|$ generate
    a terahertz carrier in a UTC-PD.
    The phase noise of the generated carrier contains the phase noise
    $S_{\phi,\mathrm{opto\text{-}THz}}(f)$ inherited from the optical frequency difference
    together with additive noise $S_{\phi,\mathrm{UTC}}^{\mathrm{add}}(f)$ introduced
    during photomixing. Residual intensity noise (RIN) of the opto-THz source is converted into phase modulation (PM) of the carrier by the UTC. Amplified spontaneous emission (ASE) from an optical amplifier (OA) also beats against the carrier to generate added noise.
    Photocurrent shot noise, Eq.~\eqref{eq:shot}, and thermal noise provide the remaining
    white contributions.
    A flicker-like ($1/f$) contribution of unidentified origin is also present.
    }
    \label{fig:noise_budget}
\end{figure}

Photomixing converts the frequency difference between two mutually coherent
optical fields into an electrical carrier. Two optical
fields at frequencies $\nu_1$ and $\nu_2$ incident on a sufficiently fast photodiode
generate a photocurrent frequency component at
\begin{equation}
    \nu_{\mathrm{THz}} = |\nu_2-\nu_1|,
\end{equation}
with a phase follows the relative phase of the two optical fields. The spectral purity of
an optical reference can therefore be transferred directly to the terahertz domain without an intermediate electronic multiplication chain.

The principle and the relevant noise processes of photomixing are summarized in
Fig.~\ref{fig:noise_budget}. The optical frequency difference carries a
phase-noise power spectral density $S_{\phi,\mathrm{opto\text{-}THz}}(f)$, which
photomixing transfers to the generated electrical carrier, while the UTC-PD introduces an
additional contribution $S_{\phi,\mathrm{UTC}}^{\mathrm{add}}(f)$. Treating these as
statistically independent, the phase-noise power spectral density of the generated carrier
is
\begin{equation}
    S_{\phi,\mathrm{tot}}(f)
    =
    S_{\phi,\mathrm{opto\text{-}THz}}(f)
    +
    S_{\phi,\mathrm{UTC}}^{\mathrm{add}}(f).
    \label{eq:budget}
\end{equation}
Independence is evident for amplifier and shot noise, which originate in processes
uncorrelated with the optical phase. It is less obvious for AM-to-PM conversion, which is
driven by intensity fluctuations of the same field that carries the reference phase. As will be described below, the AM-to-PM conversion is experimentally reduced in this work well below other sources of noise, leaving its statistical independence a moot point. The purpose of the measurements
described below is to determine the magnitude and physical origin of
$S_{\phi,\mathrm{UTC}}^{\mathrm{add}}(f)$, and we consider the five mechanisms
that contribute to it: AM-to-PM conversion, ASE-signal beating, current shot noise, thermal noise, and low-frequency flicker. The corresponding equations are used, without re-derivation, in
Sections~\ref{sec:apc}--\ref{sec:white} to convincingly interpret the measurements. Note that in this section, we quote phase noise power spectral density as $S_\phi(f)$ [rad$^2$/Hz] throughout. In the experimental part of this work, we use the more experimentally common quantity $\mathscr{L}(f) = 10 \log_{10}(S_\phi(f)/2)$ [dBc/Hz]. 

\subsection{AM-to-PM conversion}

Optical relative-intensity noise (RIN) can be converted into phase fluctuations through the
nonlinear response of a saturated photodiode. Writing the one-sided RIN power spectral density as $S_I(f)$, this contribution is
\begin{equation}
    S_{\phi,\mathrm{AM\text{-}PM}}(f) = \alpha^2 S_I(f),
    \label{eq:ampm}
\end{equation}
where the conversion coefficient $\alpha^2$ has units of rad$^2$ and depends strongly on
 UTC-PD photocurrent and reverse bias. The same RIN therefore produces
substantially different phase noise depending on the operating point of the photodiode,
and, as shown in Section~\ref{sec:apc}, $\alpha^2$ varies by more than 30~dB over the
accessible range of bias voltages. Because $S_I(f)$ inherits the spectral shape of the
RIN, this contribution is not white and is largest at low and
intermediate Fourier frequencies.

\subsection{ASE-signal beating} \label{sec:asesig}

If optical amplification precedes photodetection, beating between amplified spontaneous
emission (ASE) and the coherent optical signal introduces a white phase-noise floor
\begin{equation}
    S_{\phi,\mathrm{ASE}}(P_{\mathrm{in}})
    = \frac{2 h\nu F}{P_{\mathrm{in}}},
    \label{eq:ase}
\end{equation}
where $h$ is Planck's constant, $\nu$ the optical frequency, $F$ the amplifier noise
figure, and $P_{\mathrm{in}}$ the optical power at the amplifier input. For a typical
erbium-doped fibre amplifier with $F = 6$~dB driven at 0~dBm, Eq.~\eqref{eq:ase} gives
$-149.9$~dBrad$^2$/Hz. This exceeds the floor measured in this work by roughly 8~dB,
so optical amplification alone can set the noise floor even when the underlying optical
reference is far quieter. Eliminating amplification from the signal path is
therefore a precondition for accessing the UTC-PD floor.

\subsection{Current shot noise}

Once ASE and AM-to-PM contributions are suppressed, photocurrent shot noise provides an additive contribution \cite{Quinlan2023}. The photocurrent carries a one-sided shot-noise current spectral density $2eI$, with $e$ the elementary charge and $I$ the average photocurrent. The THz carrier is the photocurrent component at $\nu_{\mathrm{THz}}$, of peak amplitude $I_\omega$. Carrier and noise drive the same load of real impedance $R$, so that $P_{\mathrm{THz}}=I_\omega^2 R/2=|H(\nu_{\mathrm{THz}})|^2 \rho I^2 R/2$ and 
\begin{equation}
    S_{\phi,\mathrm{shot}}(I) = \frac{2eIR}{P_{\mathrm{THz}}}
    =\frac{4e}{|H(\nu_{\mathrm{THz}})|^2 \rho I}.
    \label{eq:shot}
\end{equation}
Here, $|H(\nu_{\mathrm{THz}})|^2 \leq 1$ is a carrier-frequency-dependent transfer function, including transit-time effects. $\rho \leq 1$ is a factor coming from power imbalance between the two optical lines being photomixed (see the Supplemental Material for a calculation of this effect). Losses between the junction and the waveguide attenuate carrier and noise equally and leave the right-hand side of Eq.~(\ref{eq:shot}) unchanged. For the optimal values of $|H(\nu_{\mathrm{THz}})|^2$ and $\rho$, we have the bound,
\begin{equation}
    S_{\phi,\mathrm{shot}} \geq \frac{4e}{I},
    \label{eq:shotbound}
\end{equation}
corresponding to $-161.9$~dBrad$^2$/Hz
($\mathscr{L} = -164.9$~dBc/Hz) at $I = 10$~mA, for example.

\subsection{Thermal noise}

Thermal noise associated with
finite carrier power contributes a further white floor \cite{Rubiola2009}, 

\begin{equation} \label{eq:therm}
    S_{\phi,\mathrm{therm}} = \frac{k T}{P_\mathrm{THz}},
\end{equation}

\noindent where $k$ is Boltzmann's constant, and $T$ is the temperature. The ratio
of shot to thermal noise is $2eIR/kT$, which for $I = 10$~mA, $R = 50~\Omega$, and $T = 290$~K, is approximately 16~dB;
shot noise dominates thermal noise under all conditions considered in this study, and
Eq.~\eqref{eq:shot} sets the floor that can be approached at high Fourier frequency. The
two scale differently with photocurrent, since $P_{\mathrm{THz}} \propto I^2$ in the
unsaturated regime, so that $S_{\phi,\mathrm{shot}} \propto 1/I$ while the thermal
contribution falls as $1/I^2$. Verifying the predicted photocurrent dependence is
consequently a direct test of whether a measurement has reached the shot-noise regime, and
we perform this test in Section~\ref{sec:white}. We note also that there exists photon shot noise attributed to the THz photons. However, with a carrier of 300\,GHz, this is an order of magnitude below the thermal noise.

\subsection{Flicker noise}

A flicker-like contribution, having a characteristic 1/f slope, may also be present between the reference-noise and
white-noise regimes; we show in Section~\ref{sec:white} that it lies well above the level
predicted by Eq.~\eqref{eq:ampm} and therefore has a distinct origin.

\hskip1em

The additive noise of the photomixing process must finally be distinguished from noise
contributed by the measurement system. Mixer noise, IF amplification, local-oscillator
noise, and thermal noise in the detection paths can mask
$S_{\phi,\mathrm{UTC}}^{\mathrm{add}}(f)$ without representing fluctuations of the
generated carrier. At the levels predicted by Eq.~\eqref{eq:shot}, a conventional
single-channel measurement is limited by the detection system well before reaching the photomixing floor. The experimental problem is therefore to separate the
additive phase noise of the UTC-PD from both the phase noise inherited from the opto-THz
reference and the independent noise of the detection system, which we address in the
following section.

\section{Cross-correlation measurement of photomixing noise} \label{sec:setup}

\begin{figure}[htbp]
\centering\includegraphics[width=10cm]{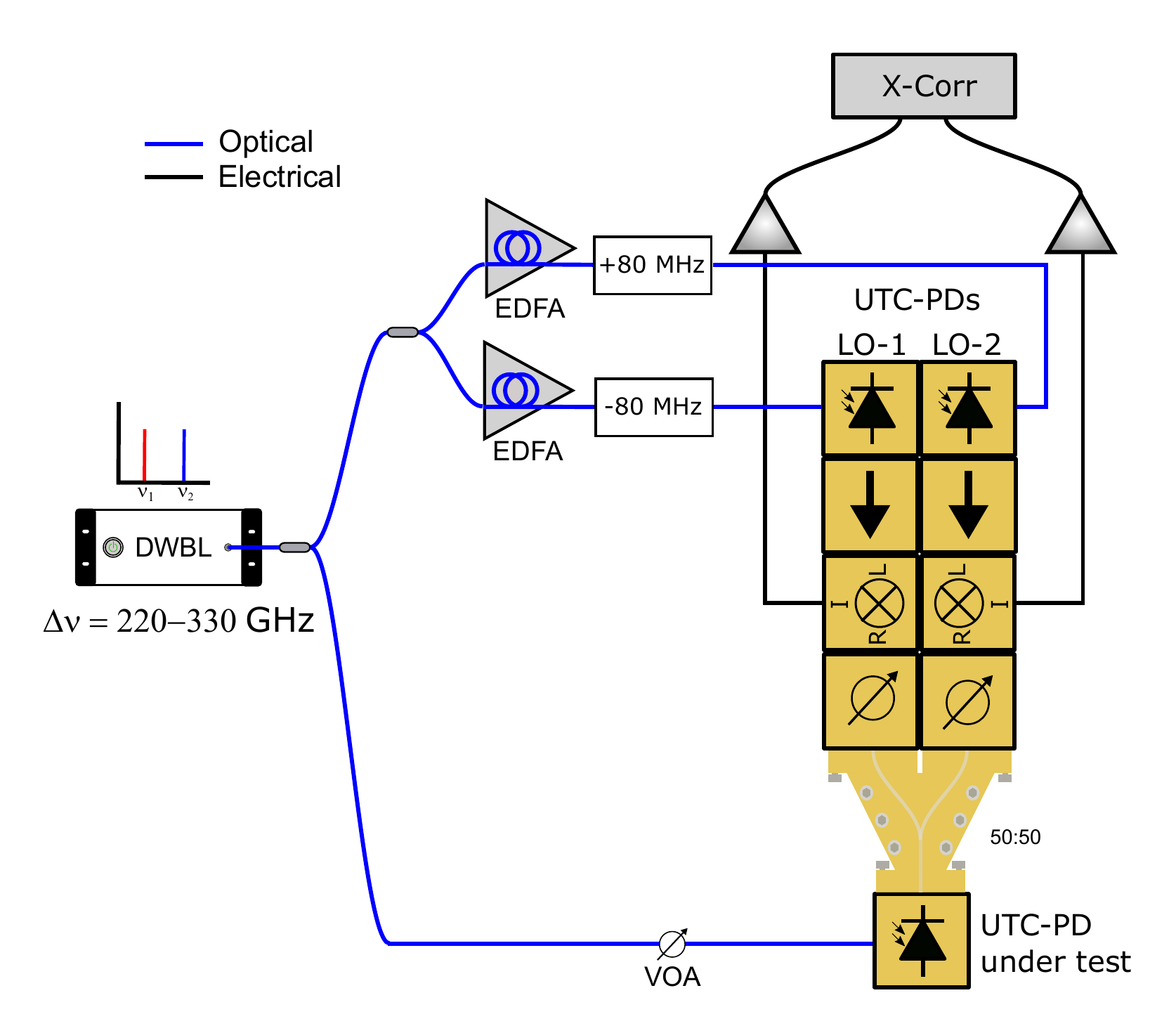}
\caption{Experimental setup for cross-correlation self-comparison measurement. A high-power DWBL was split, with most (90\%) of the power being directed toward the UTC-PD common to the two detection pathways. The rest (10\%) of the optical power was split equally in two and amplified. For each of the two paths, the optical lines were separated by DWDM, and one line was shifted by $\pm$80 MHz, with opposite order for the two paths to avoid DC interference. The lines were recombined and input to an LO UTC-PD. After passing through a THz isolator, the LO signals mixed with the common signal from the UTC-PD under test, and the 80 MHz IF was amplified and cross-correlated. THz components are shaded in gold and were WR3.4 rectangular metal waveguide coupled. EDFA: erbium-doped fiber amplifier. VOA: variable optical attenuator. X-Corr: cross-correlator.}\label{fig:setup}
\end{figure}

This work makes use of cross-correlation phase noise measurements, in which two independent measurement arms are used to average out sources of noise that are not common to the two arms. With this method, phase noise of the oscillator shared between the two arms can be measured without being limited by non-common-mode noise sources such as IF microwave thermal noise and local oscillator (LO) noise \cite{Walls1976}. Our test setup is displayed in Fig.~\ref{fig:setup}. In order to identify the noise added by photomixing, rather than measure the opto-THz reference itself, the same reference was used for the three independent UTC-PDs involved in the measurement. The WR-3.4 waveguide-coupled UTC-PDs (NTT Innovative Devices IOD-PMJ-24001) used were manufactured with SiC substrates, enabling several milliwatts of THz output power. In principle, since phase noise was shared among the three pathways, it is possible to use any opto-THz reference. Even a highly unstable one such as two freerunning diode lasers would work. However, such a setup would require careful path length matching among the three paths so as to avoid being limited by the beating between the differentially time-delayed signals \cite{Rubiola2005}. See the Supplemental Material for a calculation of this noise source, for different opto-THz references. We instead elected to use a highly stable source, a DWBL, to obviate the need for such arduous pathlength matching.

The architecture of the DWBL has been described in detail, previously \cite{Heffernan2024}. We operated the DWBL with a high output power of 80\,mW, sufficient to drive the UTC-PD without amplification. This mitigates the ASE noise discussed earlier (Section \ref{sec:asesig}). The DWBL optical output was split, with most (90\,\%) being sent to the RF detection arm. The remainder was split between the two LO arms and amplified. The two optical lines of each of these were split by a dense wavelength division multiplexer (DWDM), and one line was shifted by 80 MHz using an acousto-optic modulator (AOM). For the two arms, the AOMs were chosen to have opposite diffraction order to prevent DC interference in the cross-correlation. Each of the two optical lines were recombined, and incident on two independent UTC-PDs. The THz radiation was then used to saturate the LO ports of two Schottky barrier diode mixers. THz isolators were employed to reduce power fluctuations caused by reflections between the mixer and UTC-PDs.

For the two LO arms to remain statistically independent, it is important to avoid shared sources of noise. In order for optical amplifier noise to average out, there must be two independent erbium-doped fiber amplifiers (EDFAs) for the two paths, not a single shared one before the splitter, even if the optical power would be adequate in the latter case. The synthesizers and microwave amplifiers driving the AOMs must also be driven by separate DC power supplies for the two arms. With the same power supplies used for the two amplifiers, spurious correlation effects were observed in the cross-correlated spectrum of the experiment. These were removed by separating the power supplies.

While cross-correlation is undoubtedly a powerful technique, a statistical artifact known as cross-spectral collapse can limit its usefulness \cite{Nelson2014, Gruson2020}. Cross-spectral collapse arises due to the presence of anti-correlated noise sources --- specifically the thermal noise of the splitter impedance. This anti-correlated noise subtracts from the device-under-test phase noise to be measured, leading to spuriously low phase noise measurements, even averaging down towards zero if the two noise sources precisely cancel. The balanced mixers used in this experiment allowed a maximum RF-port power of 0.1\,mW, leading to a thermal phase noise floor of nearly -167\,dBc/Hz. Cross-spectral collapse becomes a limiting factor at measurements approaching this level \cite{Greenberg2026CLEO}.

We eliminated cross-spectral collapse by driving the UTC-PD under test at high power, generating about 1 mW of THz radiation, then attenuating the radiation after splitting between the two detection paths. In this configuration, the thermal noise at the splitter, calculated using Eq. \eqref{eq:therm}, was below -170\,dBc/Hz, an insignificant level for these measurements. Thermal noise after the splitter (including from the attenuators) was uncorrelated between the two arms and averaged away in the cross spectrum. Following the attenuators, the RF THz signal was input to the RF port of a Schottky barrier diode mixer. The IF of the two mixers was amplified (taking care that the amplifiers not be powered by the same power supply), and input to a phase noise analyzer capable of cross-correlation (Microchip 53100A). No evidence of cross-spectral collapse, as evidenced by the characteristic `notch' in the cross spectrum, was observed during this campaign. This cross-correlation setup could then be used to measure the UTC-PD noise floor.

\section{Phase noise floor at fixed carrier frequency}

\begin{figure}[htbp]
\centering\includegraphics[width=9cm]{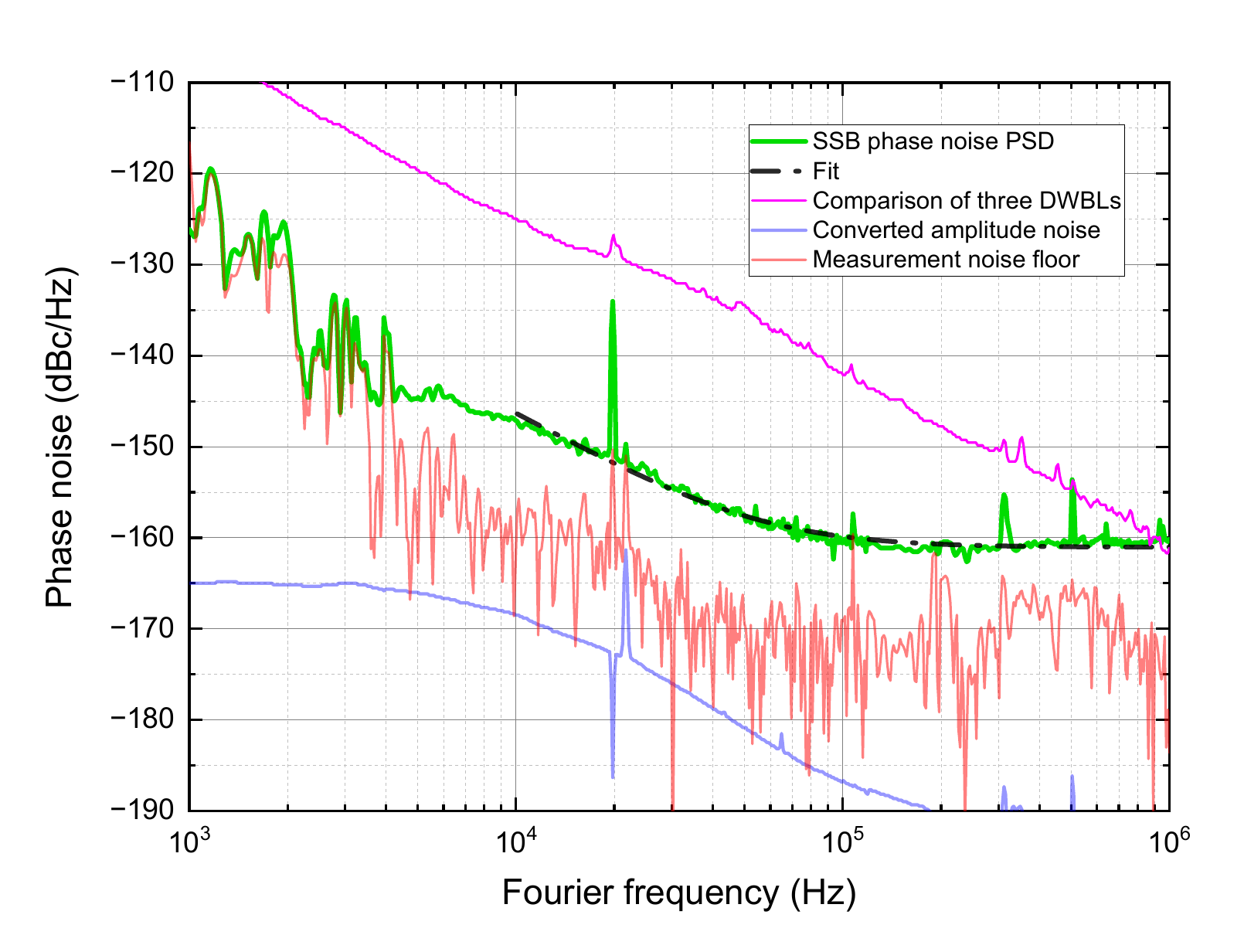}
\caption{Measured phase noise at a carrier frequency of 263.2 GHz. The three UTC-PDs are driven by 10 mA of photocurrent, and the bias voltage is adjusted to minimize the AM-to-PM coefficient for each UTC-PD independently. The red line is the phase noise analyzer floor, as estimated by the imaginary part of the cross-spectrum. The black dash-dotted line is a fit, using the heuristic power-law model described in the text (Eq. \eqref{eq:heuristic}). The small spurs at 20, 300, and 500 kHz were excluded from the fit, which was applied between the Fourier frequencies of 10 kHz and 1 MHz. In this data, the phase noise floor is found to be -161.0 dBc/Hz.  The pink trace is a comparison with three independent DWBLs, reaching a similar high Fourier-frequency phase noise level to the floor from the self-comparison. The blue trace is the measured intensity noise, scaled by the AM/PM coefficient. }\label{fig:championdata}
\end{figure}

The cross-correlation setup shown in Fig.~\ref{fig:setup} was used to investigate the noise floor of photomixing at 263.2 GHz. Performance below -160 dBc/Hz is shown in Fig.~\ref{fig:championdata}; to our knowledge, this is the lowest white phase-noise floor reported for direct photomixing in the terahertz range. For this data, the AM-to-PM conversion was minimized by tuning the bias voltage; see Section~\ref{sec:apc}. The white noise floor is obtained for the decade between 100 kHz and 1 MHz. Below a few kHz, the phase noise measurement is limited by averaging of acoustic noise from unshared fiber paths between LO-1 and LO-2, as shown by a phase-noise-analyzer floor at the level of the measured UTC-PD phase noise. A flicker-like noise source is prominent between about 5 and 100 kHz. The source of this flicker-like noise is not clear; we note it is some 20\,dB above the calculated AM-to-PM conversion after the minimization procedure described in Section~\ref{sec:apc}. As this work is concerned with noise floor rather than phase noise at lower Fourier frequency, we elect to treat this noise heuristically. This noise did not affect the white noise floor revealed by the fit. In order to accurately extract a noise floor, a fit was performed between 10 kHz and 1 MHz. The data are well described by a fit of the form,
\begin{equation}
    \mathscr{L}_\text{tot}(f) = 10 \log_{10} \left[ 10^{\mathscr{L}_0/10} + b/f^a \right],
    \label{eq:heuristic}
\end{equation}
\noindent where $\mathscr{L}_0$ is the phase noise floor in dBc/Hz, and the second term describes a power-law (flicker-like) noise process with slope $-a$ on a logarithmic scale. Several spurs were present in the phase noise spectrum, at 20, 300, and 500 kHz, and these were excluded from the fitting window. 

The noise floor measured here is supported by a comparison using three independent DWBLs (pink trace of Fig.~\ref{fig:championdata}), in which the three UTC-PDs of Fig. \ref{fig:setup} were driven by separate optical references rather than three branches of a common DWBL. The phase noise of the signal under test is common to the two mixer outputs and is therefore retained in the cross-spectrum, whereas the independent LO-source noise averages away. Though the self-comparison only measures the residual noise of photomixing, DWBLs are capable of actually accessing this noise floor, at reasonable Fourier frequencies of about 1\,MHz.

\subsection{Experimental characterization of AM-to-PM conversion} \label{sec:apc}

The photomixed microwave literature established that amplitude noise of the optical input of a diode can induce phase noise on the microwave output, through the intrinsic nonlinearity of the saturated photodiode \cite{Zhang2012, Bouchand2017}, though this effect has not yet been systematically investigated in the THz regime. To probe this effect, a voltage-controlled variable optical attenuator was placed before the UTC-PD under test in Fig.~\ref{fig:setup}, and it was driven by a function generator at 1\,kHz to induce strong amplitude modulation. The amplitude and phase noise of the feature --- in both cases, well above the background noise of the measurement --- were extracted from the cross-spectrum measured by the phase noise analyzer.

\begin{figure}[htbp]
\centering\includegraphics[width=13cm]{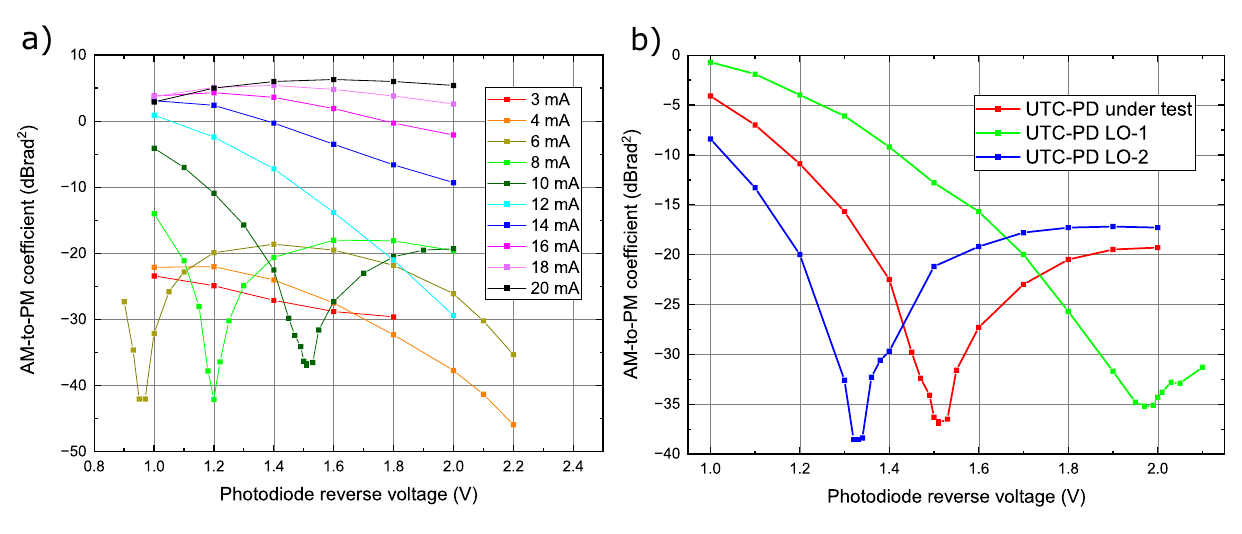}
\caption{AM-to-PM coefficient dependence on photocurrent, bias voltage, and diode, measured at a carrier frequency of 263.2 GHz. (a) Measurement at variable photocurrent and bias voltage. For a fixed photocurrent, it is often possible to adjust the bias voltage to find a minimal condition for the AM-to-PM coefficient, of -35 dB or below. As the photocurrent increases, the minimum of the coefficient shifts to a reverse bias of greater magnitude. (b) With the photocurrent fixed to 10 mA, three UTC-PDs with identical model numbers are examined, and significant variability among these nominally identical units is observed. }\label{fig:APCbycurrent}
\end{figure}

The AM-to-PM conversion coefficient $\alpha^2$, defined by Eq.~\eqref{eq:ampm}, was extracted from the measured phase noise and relative-intensity noise spectra. This coefficient was measured for the three UTC-PDs employed in this setup, and for a variety of different bias and photocurrent conditions. Fig.~\ref{fig:APCbycurrent}(a) shows $\alpha^2$ for the UTC-PD under test. It is measured for photocurrents ranging from 3 to 20 mA, and reverse-bias magnitudes ranging from 0.9 to 2.2 V. All measurements were performed at a carrier frequency of 263.2 GHz. Similar to the results of optically derived microwave signals \cite{Zhang2012}, it was found that, for a fixed photocurrent, a minimum of $\alpha^2$ can be found by tuning the bias voltage of the UTC-PD. As the photocurrent increases, a greater voltage must be applied to attain the optimal condition for $\alpha^2$. Beyond 12\,mA, a reverse voltage greater than the specified maximum bias would have to be applied, rendering the optimal working point experimentally unobtainable. Fig.~\ref{fig:APCbycurrent}(b) shows $\alpha^2$ measured with a photocurrent of 10\,mA for the three UTC-PDs used in this experiment, all made with ostensibly the same industrial processes. The three devices exhibit qualitatively similar conversion coefficients, decreasing $\alpha^2$ as the reverse voltage increases, leading to a minimum of -35 dBrad$^2$ or below, followed by gradual increase. However, the minima occur at significantly different reverse-bias magnitudes for the three UTC-PDs, ranging from 1.33 V to 1.95 V. In the measurements that follow, each of the LO UTC-PDs were operated at 10 mA of photocurrent, generating sufficient THz power to saturate the LO port of the mixer, and with a reverse voltage tuned to minimize $\alpha^2$. We note that, if it can be assumed that all of the AM noise on the LO UTC-PDs comes from their respective EDFAs, then this source of noise would be uncorrelated and average away. However, converted amplitude noise from the source DWBL would be partially correlated (or even anti-correlated) and lead to noise that would not average away in the cross spectrum. To prevent this, each of the three UTC-PDs are tuned to the AM-to-PM null condition. With $\alpha^2 < -30$\,dBrad$^2$, intensity to phase conversion was an insignificant source of noise in these measurements, across all Fourier frequencies; see the blue trace of Fig.~\ref{fig:championdata}.

\subsection{White noise from optical amplification} \label{sec:white}

\begin{figure}[htbp]
\centering\includegraphics[width=9cm]{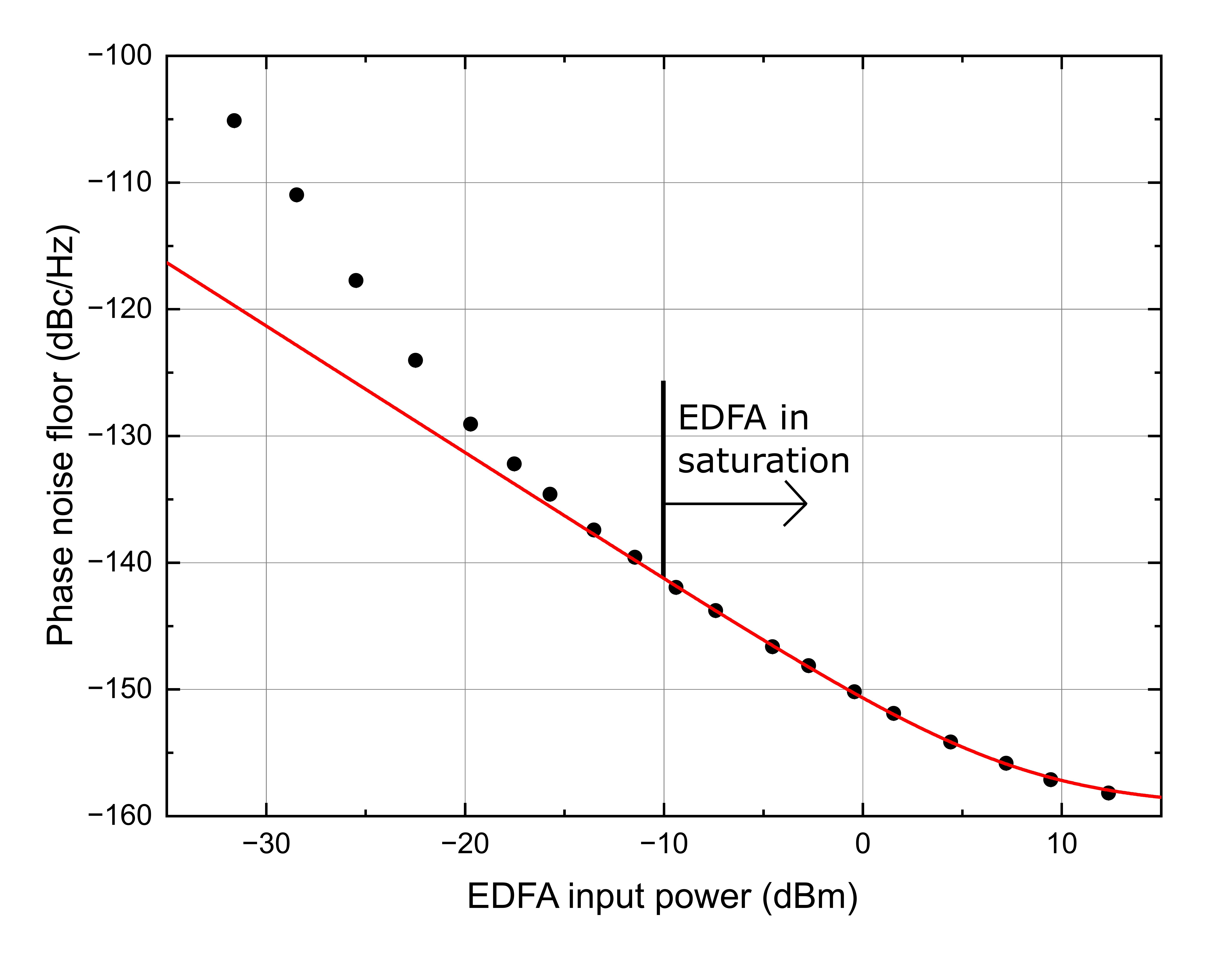}
\caption{Measured phase noise floor when an EDFA is included in the common path of the cross-correlation measurement. As the optical input power is decreased, the amount of amplified spontaneous emission (ASE) increases relative to signal, leading to an increasing noise floor. The red line is a fit to the equation in the main text. The fit is applied between -10 and +13 dBm, the range of input powers in which the EDFA, with 40 dB of gain and 300 mW of saturated power, exhibits saturated output. The specified noise figure is valid only in saturated operation, as shown by excess noise for low input power. Error bars are smaller than the symbols.}\label{fig:ase_noisefloor}
\end{figure}

Optical amplification can be a limiting noise source in this setup, as anticipated in Section~2 (see Eq.~\eqref{eq:ase}); to quantify its contribution directly, an EDFA was introduced before the UTC-PD under test, with a VOA before the EDFA allowing the input optical power to be adjusted. For each setting of optical input power, the phase noise floor was measured by fitting to Eq.~\eqref{eq:heuristic}. The UTC-PD photocurrent was held constant at 10 mA, as the EDFA input changed. The dependence of phase noise floor on optical input power is shown in Fig.~\ref{fig:ase_noisefloor}.

The white noise from beating between amplified spontaneous emission (ASE) and the carrier follows Eq.~\eqref{eq:ase} \cite{Quinlan2023}, where $P_\text{in}$ is, again,  the optical input power to the EDFA. The data are fit to an equation including an offset,
\begin{equation}
    \mathscr{L}_\text{tot}(P_\text{in}) = 10 \log_{10} \left[ 10^{\mathscr{L}_\text{ASE} (P_\text{in}) /10}+ 10^{\mathscr{L}_\text{off}/10} \right]. \label{eq:aseadd}
\end{equation}

The noise figure is specified by the manufacturer as 6\,dB in saturation. As a result, we only fit data between -10 and +13 dBm, the range at which the amplifier output is saturated. The data of Fig.~\ref{fig:ase_noisefloor} are well described by Eq.~\eqref{eq:aseadd}, and the ASE noise for an optical input of 0\,dBm amounts to -151.3\,dBc/Hz. This is close to the prediction from Eq.~\eqref{eq:ase}, $\mathscr{L}_\text{ASE} = -152.9$\,dBc/Hz. We emphasize clear evidence of excess noise figure for low optical input power, as exemplified by a noise floor significantly above the saturated-output fit. These findings demonstrate how optical amplification can introduce significant white noise to a photomixed source, and care should be taken to prevent this effect from limiting a measurement.

\begin{figure}[htbp]
\centering\includegraphics[width=9cm]{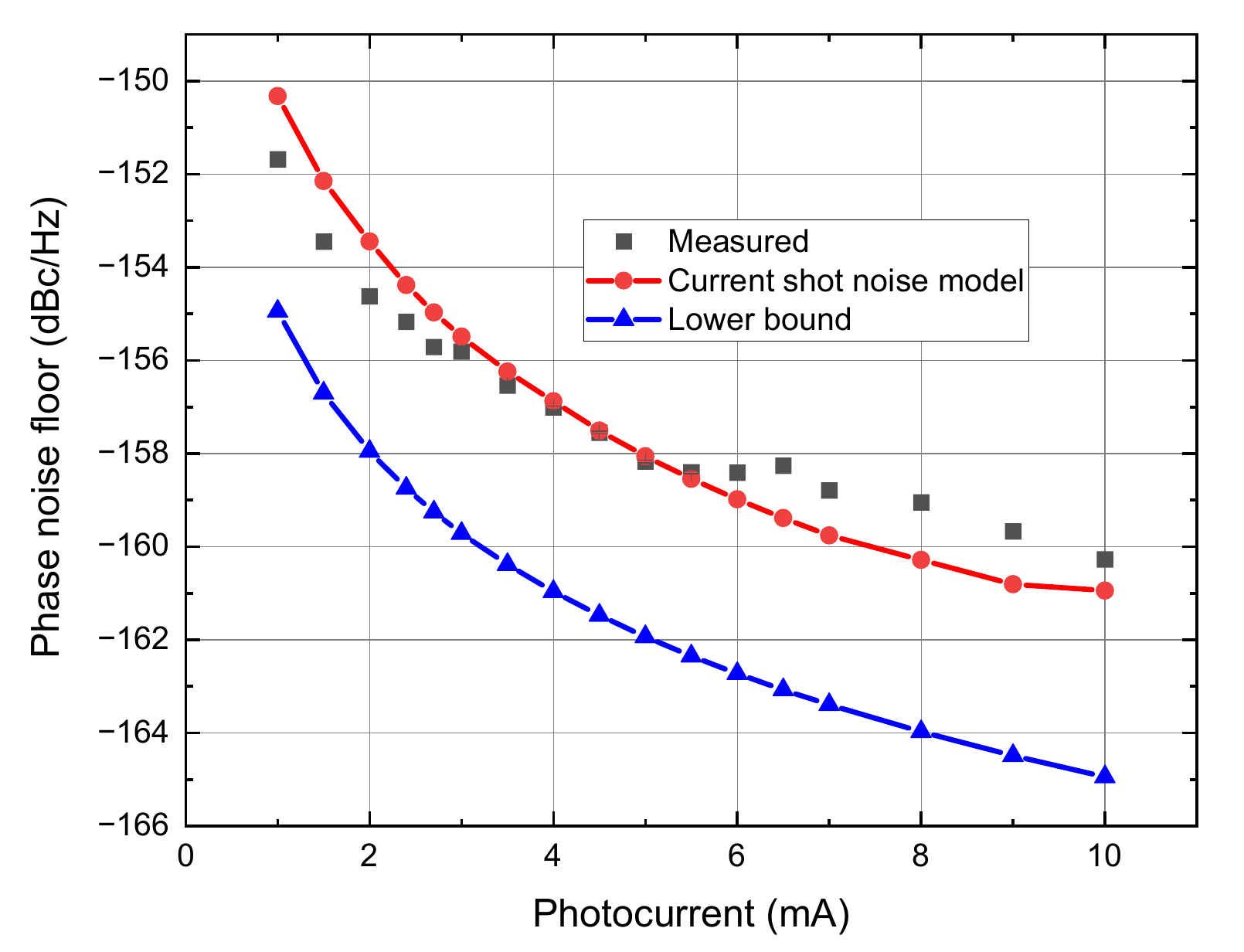}
\caption{Phase noise floor versus photocurrent at a carrier frequency of 263.2~GHz. Black squares are the measured phase noise floor as a function of photocurrent. The phase noise floor is described well by a current shot noise model (see main text). For these measurements, the bias voltage was adjusted to minimize the AM-to-PM coefficient at the respective photocurrent. The red circles are the calculated current shot noise floor, including the effects of changing bias voltage on the THz power, as measured by a calibrated THz thermal power meter. The blue triangles are a lower bound, for ideal photo-mixing (see Eq. \ref{eq:shotbound}).
}\label{fig:photocurrent_dependence}
\end{figure}

\subsection{Photocurrent shot noise}

We expect the limiting factor for white noise to be shot noise of the UTC-PD photocurrent. To test this, the photocurrent was systematically varied, and the phase noise floor was extracted under each set of conditions by fitting Eq.~\eqref{eq:heuristic}. Since $\alpha^2$ changes with photocurrent (see Fig.~\ref{fig:APCbycurrent}), it was necessary to adjust the bias voltage along with the photocurrent. Below 6 mA, the minimum of $\alpha^2$ is below the specified acceptable range of reverse bias voltages, so the reverse-bias magnitude was maintained at 1\,V. However, this was not a significant systematic, as $\alpha^2$ under these low-current, below-saturation conditions is intrinsically low. These measurements (black squares) are shown in Fig.~\ref{fig:photocurrent_dependence}.

The measurements are compared to the theoretical prediction of current shot noise, Eq.~\eqref{eq:shot} \cite{Rubiola2009}, with $P_\text{THz}$ the THz power generated by the UTC-PD. The THz power is measured by a calibrated waveguide-coupled thermal powermeter (Virginia Diodes PM5B) situated after a short length of waveguide. We note that the impedance can differ substantially from the nominal $R = 50\,\Omega$. Further, the power measured depends on the efficiency by which THz power generated by the UTC-PD is coupled into the waveguide. This in turn depends on impedance matching between the UTC-PD and the patch antenna. Assuming a resistance of $50 \, \Omega$ and ideal impedance matching, the predicted current shot noise (red circles) matches the measured phase noise floor (black squares); see Fig.~\ref{fig:photocurrent_dependence}. However, this level of absolute agreement is not necessarily expected, due to uncertainties in both impedances. On the other hand, since both impedances are nearly invariant with photocurrent, the similarity of the trends provides strong evidence that the noise floor is limited by current shot noise. For comparison, the blue triangles show the phase noise floor with ideal photomixing (see Eq. \eqref{eq:shotbound}) is calculated. The measured noise floor approaches within 6\,dB of this fundamental limit.

\section{Phase noise floor across the WR-3.4 band}

Though the above measurements were realized at a carrier of 263.2 GHz, it is desirable to check the applicability of these methods across a range of carrier frequencies, since fields like communications and radioastronomy differ greatly in the carrier frequencies utilized. For this reason, the DWBL frequency was adjusted across the whole of the WR-3.4 band, from 220 to 330 GHz. It was found during these measurements that the minimum condition for $\alpha^2$ varied with carrier frequency. However, across the range of carrier frequencies explored here, $\alpha^2$ was small enough that excess noise from amplitude to phase conversion did not compromise measurements of the noise floor, and the heuristic fit, Eq.~\eqref{eq:heuristic}, could still be employed to determine the noise floor, without minimizing $\alpha^2$ at each carrier frequency.

\begin{table}[htbp]
\caption{Phase and timing noise across the WR-3.4 band and slightly beyond. The THz power is measured with 10 mA of photocurrent and a reverse-bias magnitude of 1.5 V.}
  \label{tab:timingnoise}
  \centering
\begin{tabular}{llll}
\hline
Carrier frequency & Phase noise floor & Timing noise floor & THz power \\
(GHz) & (dBc/Hz) & (zs/$\sqrt{\text{Hz}}$) & (mW) \\
\hline
220.5 & -160.4 & 9.7 & 0.347\\
241.4 & -160.9 & 8.4 & 0.559\\
263.2 & -161.0 & 7.6 & 1.032\\
283.6 & -162.4 & 6.0 & 0.337\\
305.5 & -158.5 & 8.7 & 0.595\\
320.0 & -157.1 & 9.9 & 1.145\\
333.6 & -159.2 & 7.4 & 0.834\\
\hline
\end{tabular}
\end{table}

The phase noise at a photocurrent of 10\,mA and bias of -1.5\,V was measured across the WR-3.4 band, as shown in Table~\ref{tab:timingnoise}, along with the timing noise. The noise floor remains below 10 zs/$\sqrt{\text{Hz}}$ across the whole measured span. The THz power in each condition was measured by thermal powermeter.

From a naive application of Eq.~\eqref{eq:shot}, it would be expected that the white phase noise floor would be anti-correlated with power, assuming the floor remains limited by current shot noise. This expectation is not confirmed by the data, as can be seen by inspection of Table~\ref{tab:timingnoise}. On the other hand, it is important to note that the UTC-PD and antenna impedances change substantially as the carrier frequency changes, potentially by orders of magnitude \cite{Natrella2016, Natrella2016a, Ssali2026}. Therefore, both the UTC-PD resistance and the coupling coefficient from the UTC-PD to the waveguide are strongly variable with carrier frequency. These considerations may explain the distribution of the noise floor measurements. Detailed simulation of the UTC-PD models used in this work could confirm this explanation \cite{Natrella2016a, Ssali2026}, though such modeling is beyond the scope of this work.

\section{Discussion and conclusion}

In this work, we have adapted cross-correlation methods to measure the phase noise of an opto-THz photomixer, minimizing cross-spectral collapse by attenuating the terahertz signal after the splitter. We have identified and experimentally separated the principal noise mechanisms that limit direct optical-to-THz photomixing with high-power UTC-PDs. By controlling AM-to-PM conversion through the photodiode current and bias and eliminating optical amplification from the signal path, the white-noise floor of the photomixing process was identified. At 263.2~GHz, we measure a white phase-noise floor of $-161.0$~dBc/Hz, corresponding to a timing-noise floor of 7.6~zs/$\sqrt{\mathrm{Hz}}$. A comparison using three independent DWBLs reaches a similar noise floor, providing an independent confirmation that the result is not an artifact of the self-comparison measurement and is experimentally relevant for a real opto-THz source.

The dependence of the measured white-noise floor on photocurrent is consistent with the scaling expected from photocurrent shot noise. The absolute shot-noise prediction depends on the real impedance of the UTC-PD and on the coupling efficiency between the photodiode, antenna, and waveguide, quantities that are not independently determined here.

This performance is not restricted to a single carrier frequency: across the WR-3.4 band, from 220 to 330~GHz, we measure timing-noise floors below 10~zs/$\sqrt{\mathrm{Hz}}$, with a minimum of 6.0~zs/$\sqrt{\mathrm{Hz}}$ at 283.6~GHz. The scatter across the measured span is consistent with the frequency-dependent UTC-PD and antenna-coupling impedance discussed above.

The measurements also reveal a flicker-like contribution between approximately 10 and 100~kHz that remains unexplained. After optimization of the UTC-PD operating point, this contribution lies well above the phase noise predicted from measured RIN and AM-to-PM conversion, indicating that another technical or intrinsic noise process is present, such as flicker noise of the UTC-PD. Although it does not limit the white-noise floor investigated here, identifying its origin will be important for applications requiring similarly low phase noise at smaller Fourier frequencies.

Taken together, these results show that direct photomixing can transfer optical spectral purity to the terahertz domain with a white timing-noise floor below 10~zs/$\sqrt{\mathrm{Hz}}$ over more than 100~GHz of carrier-frequency span. In contrast to electronic multiplication, which carries the phase fluctuations of a microwave reference to higher carrier frequencies and accumulates additional noise through successive multiplication and amplification stages, photomixing performs the optical-to-THz conversion in a single step. Once AM-to-PM conversion and optical-amplifier noise are suppressed, the achievable white-noise floor can approach that imposed by the photocurrent itself. This establishes high-power direct photomixing as a route to ultralow-noise terahertz carriers for coherent communications, spectroscopy, and precision measurement.

Further reductions in timing noise floor should be possible as UTC-PDs support larger photocurrents and higher output powers. Pulsed photodetection may also provide a route to ultra-low phase noise by exploiting correlations between shot noise and the detected optical waveform which can reduce the effective phase-noise contribution \cite{Quinlan2013}. Combining these photodetection methods with optical frequency division, including Kerr optical frequency division \cite{Egbert2026}, may therefore extend the spectral purity available from optical references into the terahertz domain at still lower timing-noise levels.

\section*{Acknowledgments}

We acknowledge Hideyuki Ohtake and Yuki Ichikawa for their support throughout this work.

%%%%%%%%%%%%%%%%%%%%%%% References %%%%%%%%%%%%%%%%%%%%%%%%%
%%%%%%%%%% If using BibTeX:
\bibliography{sample}
%%%%%%%%%% If preparing manually:
% \begin{thebibliography}{1}
% \newcommand{\enquote}[1]{``#1''}
% \bibitem{Zhang:14}
% Y.~Zhang, S.~Qiao, L.~Sun, Q.~W. Shi, W.~Huang, L.~Li, and Z.~Yang,
%   \enquote{Photoinduced active terahertz metamaterials with nanostructured
%   vanadium dioxide film deposited by sol-gel method,}
%   {\protect\JournalTitle{Optics Express}} \textbf{22}, 11070--11078 (2014).
% \bibitem{Optica}
% {Optica}, \enquote{{Optica Publishing Group},}
%   \url{http://www.opg.optica.org}.
% \bibitem{FORSTER2007}
% P.~Forster, V.~Ramaswamy, P.~Artaxo, T.~Bernsten, R.~Betts, D.~Fahey,
%   J.~Haywood, J.~Lean, D.~Lowe, G.~Myhre, J.~Nganga, R.~Prinn, G.~Raga,
%   M.~Schulz, and R.~V. Dorland, \enquote{Changes in atmospheric consituents and
%   in radiative forcing,} in \enquote{Climate Change 2007: The Physical Science
%   Basis. Contribution of Working Group 1 to the Fourth Assesment Report of
%   Intergovernmental Panel on Climate Change,}  S.~Solomon, D.~Qin, M.~Manning,
%   Z.~Chen, M.~Marquis, K.~B. Averyt, M.~Tignor, and H.~L. Miler, eds.
%   (Cambridge University Press, 2007).
% \end{thebibliography}
\end{document}

\maketitle

% \section{Noise floor measurement with three independent oscillators}

% Measurements in the main text are performed through a cross-correlated self-comparison. This is advantageous, as the phase noise of the shared reference is common to the three arms and is not limiting for measurements of the UTC-PD phase noise, the quantity of interest in this study. We emphasize that the shared reference oscillator still permits measurement of the photodiode noise floor. To demonstrate this, we perform a proof-of-concept measurement with three separate DWBLs. In this measurement, the RF DWBL is run in a high-power mode to avoid needing amplification. As shown in Fig. \ref{fig:threeDWBL}, while the phase noise is much higher compared to the self-comparison results in the main text, a noise floor below -160 dBc/Hz has been measured at 1 MHz, in agreement with the noise floor assessment reported in the main text. We emphasize that the superior performance of the DWBL is what allows this measurement, and a less stable opto-THz reference would remain several 10 dB above the noise floor, even for a Fourier frequency of 1 MHz.

% \begin{figure}[htbp]
% \centering
% \fbox{\includegraphics[width=0.8\linewidth]{figs/ThreeDWBLComparison.pdf}}
% \caption{The green line is a self-comparison of DWBLs, as described in the text. The magenta line uses three independent DWBLs, each driving one of the UTC-PDs in Fig. 1.}
% \label{fig:threeDWBL}
% \end{figure}

\section{Delay interference of oscillator noise}

When an oscillator of phase noise PSD $S_{\phi_i}(f)$ is split, and one path is delayed by $\tau$, the beatnote for the self-comparison, limited by two-wavelength delayed interferometry (TWDI), amounts to \cite{Rubiola2005},

\begin{equation}
    S_{\phi_o}(f) = |H_\phi(jf)|^2 S_{\phi_i}(f),
\end{equation}

\noindent where the transfer function amounts to,

\begin{equation}
    |H_\phi(jf)|^2 = 4 \sin^2 (\pi f \tau).
\end{equation}

\begin{figure}[htbp]
\centering
\fbox{\includegraphics[width=0.8\linewidth]{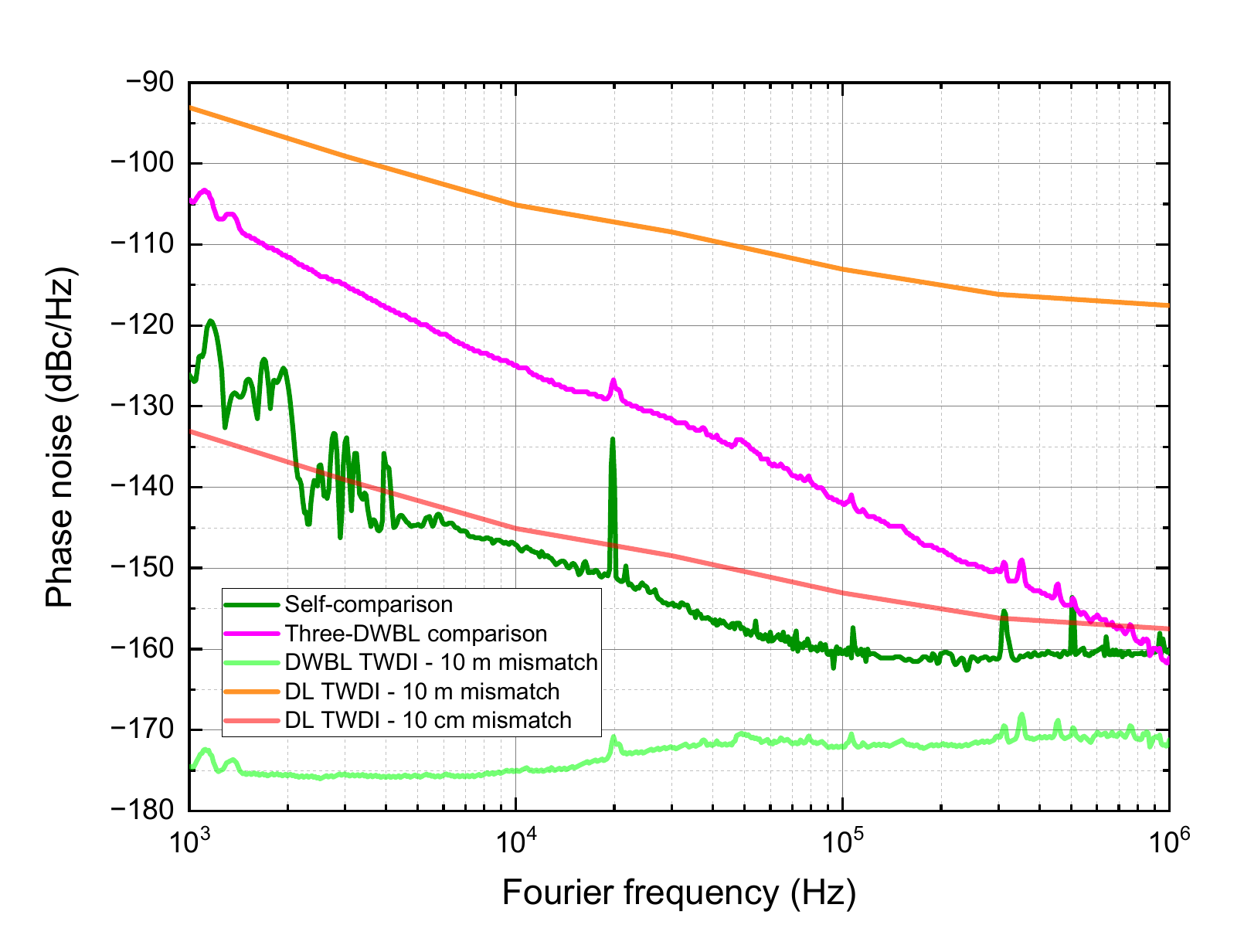}}
\caption{The green line is a self-comparison of DWBLs, as described in the text. The magenta line uses three independent DWBLs, each driving one of the UTC-PDs in Fig. 1. The light green line is the TWDI effect for a 10 m mismatch of fiber length, for a self-comparison using a DWBL. The orange and red lines are a calculation of the TWDI effects for a self-comparison using a narrow-linewidth diode laser (DL), with mismatch of 10 m and 10 cm, respectively.}
\label{fig:twdi}
\end{figure}

Assuming, conservatively, a difference of fiber length of 10 m between the RF and LO paths, the TWDI noise is $>10$\,dB below the measured phase noise (see the light green trace of Fig. \ref{fig:twdi}). Conversely, we compare against an opto-THz carrier consisting of two premium, narrow-linewidth lasers (Thorlabs ULN15K). With a 10-m imbalance, delay effects are limiting by around 50\,dB above the measured phase noise. Even with path length matching at the 10 cm level, the diode laser delay would remain the limiting factor in the phase noise measurement. Therefore, cm-scale path length matching would be necessary. This is certainly feasible, though it adds significant experimental complexity to the measurement.

It is worth noting that path length matching is necessary not only between the RF and LO paths, but also between the two LO paths used for the cross-correlation experiment. We derive this effect using the definition of the cross spectrum \cite{Nelson2014}. Allowing the delay for LO ports 1 and 2 to be $\tau$ and $\tau +\Delta\tau$, the transfer function for the cross-spectrum amounts to,

\begin{equation}
    H_\tau(jf) H_{\tau+\Delta\tau}(jf)^* = 4 \sin(\pi f \tau) \sin\left(\pi f (\tau+\Delta\tau) \right) \left[ \cos(\pi f \Delta t) + j \sin(\pi f \Delta t) \right].
\end{equation}

\noindent Here we see that, in contrast to the usual case of cross-correlation, with noise sources only correlated, uncorrelated, or anti-correlated, that the presence of a differential time delay allows the calculated cross-spectrum to have imaginary components that do not average to zero.

\section{Optical line imbalance}

If the two lines input to a photomixer have asymmetric intensities, the amplitude of the generated beatnote will be reduced. In this section, we assume the power ratio of the two lines, as measured for instance by an optical spectrum analyzer, to be $\theta = P_1/P_2$, and we will examine the generated power as a function of $\theta$, with the total optical power (and hence also photocurrent) held fixed. The electric field of the optical signal is,

% \begin{equation}
%     E(t) = A(\theta) \left[ \theta^\frac{1}{4} E_0 e^{i\nu_1 t} + \theta^{-\frac{1}{4}} E_0 e^{i\nu_2 t + \psi} \right],
%     \label{eq:eltheta}
% \end{equation}

% \noindent where $A(\theta)$ is a normalization constant to keep the total power constant, $E_0$ is an electric field amplitude, $\nu_{1(2)}$ is the frequency of line 1 (2), and $\psi$ is an arbitrary phase difference between the two lines. The normalization conditions amounts to,

% \begin{equation}
%     A(\theta) = \left(\frac{\theta^\frac{1}{2} + \theta^{-\frac{1}{2}}}{2} \right)^{-\frac{1}{2}}.
% \end{equation}

% \noindent The intensity of the photomixed wave scales as the square of the product of the first and second terms of \eqref{eq:eltheta}. The power ratio between photomixed signals from a line difference of $\theta$ and from balanced lines amounts to the quantity $\eta$,

% \begin{equation}
%     \eta = A(\theta)^4 = \left(\frac{\theta^\frac{1}{2} + \theta^{-\frac{1}{2}}}{2} \right)^{-2}.
% \end{equation}

% In this work, the optical lines are balanced to better than 2\,dB, corresponding to $\theta \geq 0.63$. This implies a power reduction factor of $\eta \geq 0.94$. So this effect is only relevant at the few percent level of precision.

\begin{equation}
    E(t) = A(\theta) \mathrm{Re}\left[\theta^{\frac{1}{4}}E_0 e^{2\pi i\nu_1 t} + \theta^{-\frac{1}{4}}E_0 e^{2\pi i\nu_2t +i\psi} \right],
\end{equation}

\noindent where $A(\theta)$ is a normalization constant to keep the total power constant as $\theta$ varies, $E_0$ is an electric field amplitude, $\nu_{1(2)}$ is the frequency of line 1 (2), and $\psi$ is an arbitrary phase difference between the two lines. Neglecting fast oscillating terms that average to zero, the optical power is,

\begin{equation}
        P(t) = \alpha |E(t)|^2 = \alpha A(\theta)^2 \left[\theta^\frac{1}{2} E_0^2 + \theta^{-\frac{1}{2}}E_0^2 + 2E_0^2 \cos{\left( 2\pi (\nu_2-\nu_1)t + \psi\right)} \right],
        \label{eq:optpower}
\end{equation}

\noindent where $\alpha$ is a geometric factor including the size of the optical beam on the photodiode. We note that the ratio of the first and second terms amounts to $\theta$, as desired. Since the last term of \eqref{eq:optpower} averages to zero over time, it is clear that the normalization factor is,

\begin{equation}
    A(\theta) = \left(\theta^\frac{1}{2} + \theta^{-\frac{1}{2}} \right)^{-\frac{1}{2}}.
\end{equation}

\noindent The photocurrent is,

\begin{equation}
    i(t) = \eta \frac{e}{h\nu} P(t) = \frac{\alpha\eta e A(\theta)^2}{h\nu} \left[\theta^\frac{1}{2} E_0^2 + \theta^{-\frac{1}{2}}E_0^2 + 2E_0^2 \cos{\left( 2\pi (\nu_2-\nu_1)t + \psi\right)} \right],
\end{equation}

\noindent where $\eta$ is the quantum efficiency of the photodiode, $e$ is fundamental charge, and $h\nu$ is the optical photon energy. The photocurrent can be decomposed into an average photocurrent, $I_\mathrm{avg}$, and a photocurrent at frequency $\nu_2-\nu_1$, with amplitude $I_\omega$:

\begin{equation}
    I_\mathrm{avg} = \frac{\alpha\eta e E_0^2}{h\nu},
\end{equation}

\begin{equation}
    I_\omega = \frac{\alpha\eta e E_0^2}{h\nu} \frac{2}{\theta^\frac{1}{2} + \theta^{-\frac{1}{2}}} = I_\mathrm{avg} \frac{2}{\theta^\frac{1}{2} + \theta^{-\frac{1}{2}}}.
\end{equation}

\noindent The THz power is

\begin{equation}
    P_\mathrm{THz} = |H(\nu_{\mathrm{THz}})|^2I_\omega^2 R/2 = |H(\nu_{\mathrm{THz}})|^2I_\mathrm{avg}^2 R/2 \left[ \frac{2}{\theta^\frac{1}{2} + \theta^{-\frac{1}{2}}}\right]^2,
\end{equation}

\noindent where $|H(\nu_{\mathrm{THz}})|^2$ is the photodiode transfer function and $R$ is the UTC-PD series resistance. This leads us to our desired result. A line imbalance of $\theta$ leads to a reduction of THz power by a factor,

\begin{equation}
    \rho = \left[ \frac{2}{\theta^\frac{1}{2} + \theta^{-\frac{1}{2}}}\right]^2.
\end{equation}

In this work, the optical lines are balanced to better than 2\,dB, corresponding to $\theta \geq 0.63$. This implies a power reduction factor of $\rho \geq 0.94$. This effect is only relevant at the few percent level of precision.

% Bibliography
\bibliography{sample}